# Normative Cerebral Perfusion Across the Lifespan

Xinglin Zeng[1], Yiran Li[1], Lin Hua[2], Ruoxi Lu[1], Lucas Lemos Franco[1], Peter Kochunov[3], Shuo Chen[4], John A Detre[5], Ze Wang[1*]

[1] Department of Diagnostic Radiology and Nuclear Medicine, University of Maryland School of Medicine, Baltimore, Maryland, United States

[2] Department of Radiology, Athinoula A. Martinos Center for Biomedical Imaging, Massachusetts General Hospital and Harvard Medical School, Boston, Massachusetts, United States

[3] Department of Psychiatry and Behavioral Science, University of Texas Health Science, SanAntonio, Texas, USA

[4] Department of Psychiatry, University of Maryland School of Medicine, Baltimore, Maryland, United States

[5] Department of Radiology, University of Pennsylvania, Philadelphia, Pennsylvania, United States

\* Corresponding authors:

Ze Wang

Corresponding author's address: Department of Diagnostic Radiology and Nuclear Medicine, University of Maryland School of Medicine, 670 W Baltimore St, HSF III, Baltimore, MD 21202, United States.

Corresponding author's phone and fax: Tel: +1 410-706-2797

Corresponding author's Email: ze.wang@som.umaryland,edu

**Abstract**

Cerebral perfusion plays a crucial role in maintaining brain function and is tightly coupled with neuronal activity. While previous studies have examined cerebral perfusion trajectories across development and aging, precise characterization of its lifespan dynamics has been limited by small sample sizes and methodological inconsistencies. In this study, we construct the first comprehensive normative model of cerebral perfusion across the human lifespan (birth to 85 years) using a large multi-site dataset of over 12,000 high-quality arterial spin labeling (ASL) MRI scans. Leveraging generalized additive models for location, scale, and shape (GAMLSS), we mapped nonlinear growth trajectories of cerebral perfusion at global, network, and regional levels. We observed a rapid postnatal increase in cerebral perfusion, peaking at approximately 7.1 years, followed by a gradual decline into adulthood. Sex differences were evident, with distinct regional maturation patterns rather than uniform differences across all brain regions. Beyond normative modeling, we quantified individual deviations from expected CBF patterns in neurodegenerative and psychiatric conditions, identifying disease-specific perfusion abnormalities across four brain disorders. Using longitudinal data, we established typical and atypical cerebral perfusion trajectories, highlighting the prognostic value of perfusion-based biomarkers for detecting disease progression. Our findings provide a robust normative framework for cerebral perfusion, facilitating precise characterization of brain health across the lifespan and enhancing the early identification of neurovascular dysfunction in clinical populations.

**Introduction**

Cerebral perfusion, also referred to as cerebral blood flow (CBF), represents the delivery of oxygen and nutrients to brain tissue through the capillary bed, a process essential for sustaining normal brain function and overall brain health (Churchill et al., 2023; Fantini et al., 2016). CBF is tightly coupled with neuronal activity through neurovascular coupling, wherein activated brain regions signal the neurovascular unit to increase blood flow, ensuring an adequate supply of oxygen and nutrients to meet and the metabolic demands (Kaplan et al., 2020; Kastrup et al., 2002; Kisler et al., 2017; Logothetis & Wandell, 2004; Schaeffer & Iadecola, 2021). Beyond its immediate role in neuronal function, CBF dynamics are closely associated with structural brain integrity, underscoring the critical role of vascular health in maintaining both brain morphology and cognitive function (Chen et al., 2013; Li et al., 2023; Ngo et al., 2024). Disruptions in CBF have been implicated in a wide range of neurological and psychiatric disorders, including stroke (Hernandez et al., 2012), Alzheimer's disease (AD) (Alsop et al., 2008; Alsop et al., 2010; Alsop et al., 2000; Camargo et al., 2021; Camargo et al., 2023; Clark et al., 2017; van Dinther et al., 2024; Wang, 2014; Wang et al., 2013; Xu et al., 2010), frontotemporal dementia (FTD) (Hu et al., 2010b; Mutsaerts et al., 2019), and major depressive disorder (MDD) (Chiappelli et al., 2023; Liao et al., 2017; Wei et al., 2018).

Previous studies have characterized the trajectories of cerebral perfusion across development and aging, consistently reporting an initial rapid increase during early childhood, stabilization around preschool years, and a gradual decline with advancing age (Biagi et al., 2007; Carsin-Vu et al., 2018; Leung et al., 2016; Ouyang et al., 2024; Paniukov et al., 2020; Wang, Fernandez-Seara, et al., 2008; Zhang et al., 2018). However, pinpointing the exact age at which CBF peaks remains challenging due to two key limitations. First, many studies have focused on restricted developmental periods with narrow age intervals, leading to findings that either show a consistent positive or negative correlation between CBF and age (Biagi et al., 2007; Leung et al., 2016; Paniukov et al., 2020). Second, while some studies encompass broader age ranges spanning childhood to late adulthood, their findings remain inconsistent and constrained by limited sample sizes (Carsin-Vu et al., 2018; Takahashi et al., 1999; Wu et al., 2016). Consequently, a comprehensive understanding of continuous lifespan dynamics in CBF, from gestation through old age, remains incomplete.

A promising approach to addressing these gaps involves the development of normative models that systematically map brain maturation across the lifespan. For instance, Bethlehem et al. (2022) constructed a normative model of brain morphometry by aggregating the largest multisite structural magnetic resonance imaging (MRI) dataset to date, offering a reproducible and generalizable framework for brain growth charts. This model utilized generalized additive models for location, scale, and shape (GAMLSS), a robust and flexible statistical framework endorsed by the World Health Organization for modeling non-linear growth trajectories (Borghi et al., 2006; Stasinopoulos & Rigby, 2008). Extending this approach to establish a cerebral perfusion chart could provide a valuable framework for elucidating CBF dynamics and their implications for brain health and disease.

Beyond lifespan characterization, normative models offer critical insights at the individual level, enabling the identification of deviations from median CBF values. Such models can serve as biomarkers for brain development and for neurological and psychiatric disorders, enhancing diagnostic precision in those conditions (Holz et al., 2023; Marquand et al., 2019; Segal et al., 2023). Given the high heterogeneity of these conditions (Segal et al., 2023; Verdi et al., 2024; Wolfers et al., 2018), normative modeling represents a powerful tool to bridge group-level biomarkers with individual-level metrics. By enabling statistical inferences for individual deviations, these models facilitate personalized diagnoses and targeted interventions (Bethlehem et al., 2022; Marquand et al., 2019; Rutherford et al., 2023; Rutherford et al., 2022). Although normative deviation scores have been applied to structural MRI and resting-state functional MRI data (Bethlehem et al., 2022; Segal et al., 2023; Sun et al., 2024; Verdi et al., 2023; Verdi et al., 2024), their potential in CBF research remains largely unexplored, despite CBF's critical role in brain function and disease mechanisms.

To address this gap, we assembled a large, multi-site neuroimaging dataset with rigorous quality control, incorporating both cross-sectional and longitudinal resting-state arterial spin labeling (ASL) perfusion MRI data. ASL MRI is the only non-invasive technique to quantify regional CBF (Detre et al., 2012; Tsujikawa et al., 2016; Wong et al., 2014). Because ASL MRI does need any exogenous contrast agents or radioactive tracers and can be repeated many times if needed, it is highly appealing to lifespan research, offering detailed insights into the dynamics of cerebral perfusion and their implications for brain health and disease across the human lifespan (Detre et al., 2012; Detre et al., 2009; Z. Wang, 2022). We first leveraged cross-sectional ASL data to construct comprehensive normative models capturing the nonlinear trajectory of cerebral perfusion at global, network, and regional levels. By integrating these models with structural MRI data, we established benchmarks for key developmental milestones. Furthermore, we assessed the clinical utility of CBF-based normative models by computing individual deviation scores relative to the 50th percentile. These scores were used to characterize disease heterogeneity in patients with AD, FTD, mild cognitive impairment (MCI), and MDD. Additionally, by integrating longitudinal ASL data, we identified typical and atypical disease progression trajectories in MCI and AD.

Our findings highlight the clinical potential of cerebral perfusion-based normative models in identifying atypical brain development, monitoring disease progression, stratifying patient subtypes, and evaluating therapeutic responses, which may potentially lead to precise and personalized diagnosis and treatment. As the first study to establish normative models of cerebral perfusion, this work provides a robust framework for individualized assessment, offering a suite of open science resources to advance standardized, quantitative evaluations of CBF. This approach fosters novel research directions into the role of vascular health in brain function and disease.

**Results**

**Mapping the normative growth of the global cerebral perfusion across the lifespan**

We aggregated multi-modal MRI data from 12,633 participants, including structural MRI and resting-state ASL MRI. After applying stringent quality control based on the quality evaluation index (QEI) and restricting the age range to birth to 85 years, the final dataset comprised 9,363 participants with high-quality imaging data. This cohort included 8,460 healthy individuals and 903 patients pooled from 20 datasets. Detailed demographics and acquisition parameters for each dataset are provided in Supplementary Tables 1 and 2.

To examine the developmental and aging trajectories of global cerebral perfusion, we characterized the normative growth patterns of its mean and variability (estimated via bootstrapping) across the lifespan (Figure 1, Panels A–D). Both shared and region-specific patterns emerged across cortical, white matter, and subcortical regions.

Global CBF exhibited a sharp increase postnatally, peaking at approximately 7.1 years (95% CI: 6.5–7.5), followed by a gradual, near-linear decline through adulthood and into later life. Cortical CBF showed the highest peak during early childhood, with substantial between-subject variability that stabilized in adulthood (Figure 1, Panels A and D). In contrast, white matter CBF followed a flatter trajectory, peaking earlier at 5.5 years (95% CI: 4.5–6.2) and declining more slowly (Panel B). Subcortical perfusion presented an intermediate pattern, peaking in late childhood to early adolescence at 6.7 years (95% CI: 6.1–7.2) and declining at a rate comparable to cortical CBF (Figure 1, Panels B and C).

Growth rate analyses (Figure 1, Panel C) further highlighted region-specific dynamics, with the sharpest perfusion increases occurring during early childhood. These findings underscore the differential timing of peak perfusion and subsequent declines across brain regions, offering critical insights into lifespan trajectories of cerebral perfusion and their potential implications for neurodevelopmental and aging-related processes.

Sex was found to significantly influence cerebral perfusion. Using GAMLSS modeling, we incorporated sex as a covariate to establish normative growth curves across the lifespan. The analysis revealed that global cerebral perfusion, including cortical, white matter, and subcortical regions, was significantly higher in females than in males ($p < 0.0001$). A significant sex difference in variance was observed only in white matter ($p_{FDR} = 0.002$), whereas no significant variance differences were detected in cortical or subcortical regions. Detailed sex-specific patterns of cerebral perfusion at the network and regional levels are listed in Supplementary Tables 3.

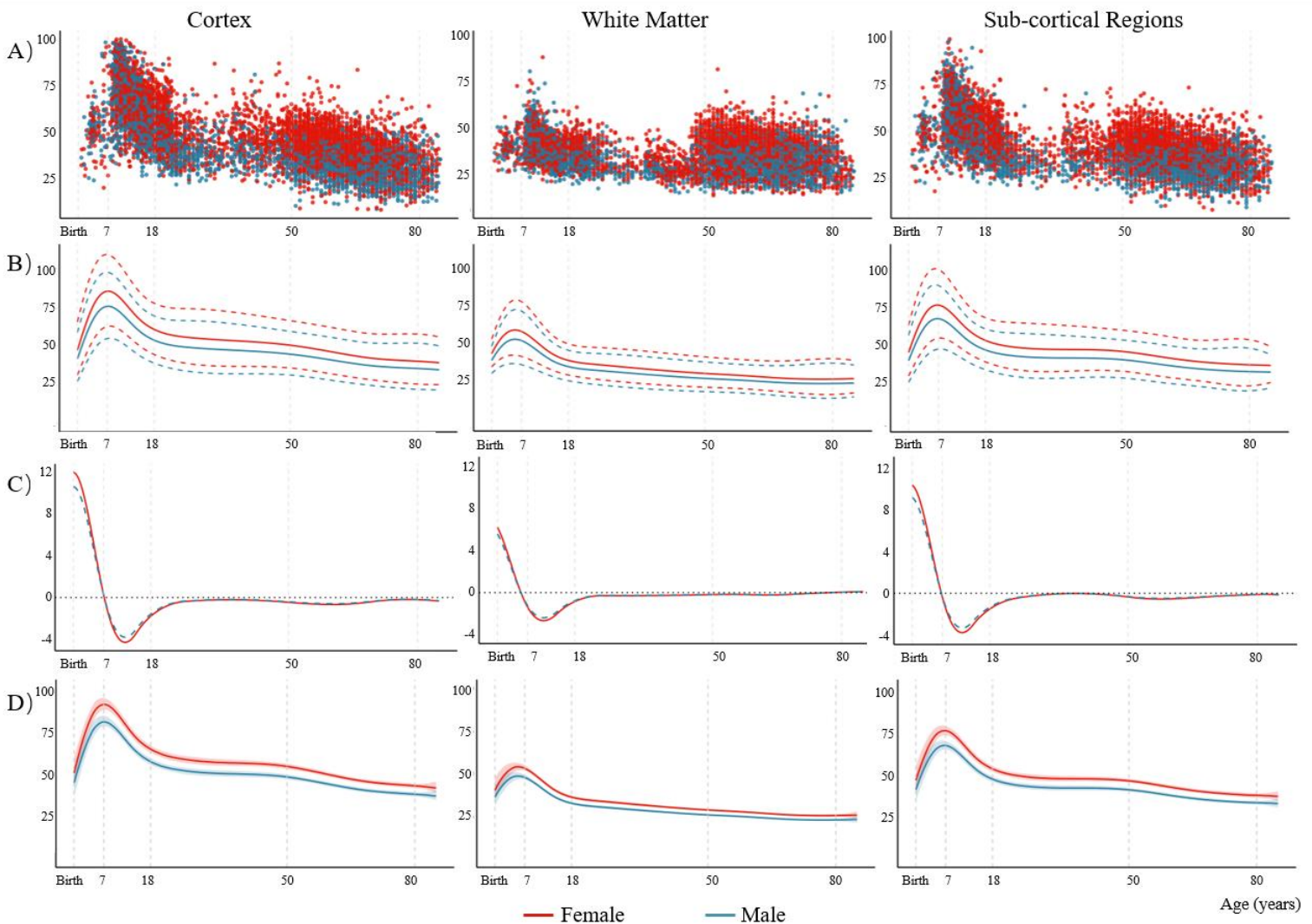

Figure 1. **Normative Cerebral Perfusion Chart.** Panel A, Raw CBF values across the cortex, white matter, and sub-cortical regions, normalized to the maximum cortical CBF. Panel B, Normative CBF Trajectories for males and females, modeled using a generalized additive model. Solid lines represent mean CBF, while dashed lines denote the 95% confidence intervals (CI). Panel C, CBF Growth Rate Across Age: illustrating a sharp decrease in early childhood, stabilization during adulthood, and a gradual decline in later life. The horizontal line (y = 0) marks the transition point where CBF shifts from increasing to decreasing, while the vertical line denotes the age of peak CBF growth. Panel D, Between-Subject Variability in CBF: Trajectories of median CBF variability and corresponding 95% CI, estimated using sex-stratified bootstrapping.

**Developmental milestones**

Key neuroimaging milestones were defined by the inflection points observed in brain trajectories (Figure 2). Global cortical cerebral perfusion peaked at approximately 7.1 years (95% CI: 6.5–7.5), closely aligning with the peak age for global gray matter volume (GMV) at 7.8 years (95% CI: 7.3–8.3). In contrast, white matter volume (WMV) peaked significantly later, around 32.5 years, consistent with its prolonged maturation trajectory. Meanwhile, cerebrospinal fluid (CSF) volume showed a continuous increase across the lifespan, reflecting the progressive ventricular expansion and brain atrophy during aging.

To extend brain charts beyond global cerebral perfusion, we applied the GAMLSS modeling approach to estimate normative regional cerebral perfusion trajectories. The analysis results revealed a distinct maturation gradient across brain regions, mirroring the hierarchical cortical gradient observed in prior neurodevelopmental studies. This maturation followed a structured sequence, initiating in sensorimotor areas, progressing through the limbic system, posterior temporal cortex, parietal cortex, and eventually reaching the prefrontal cortex, with peak ages ranging from 6.2 to 8.6 years (highlighted in the gray circle in Figure 2). Additionally, we examined relative perfusion retention at key lifespan benchmarks—18, 50, and 80 years—highlighting characteristic age-related declines in metabolic and vascular demands (illustrated in the bottom section of Figure 2).

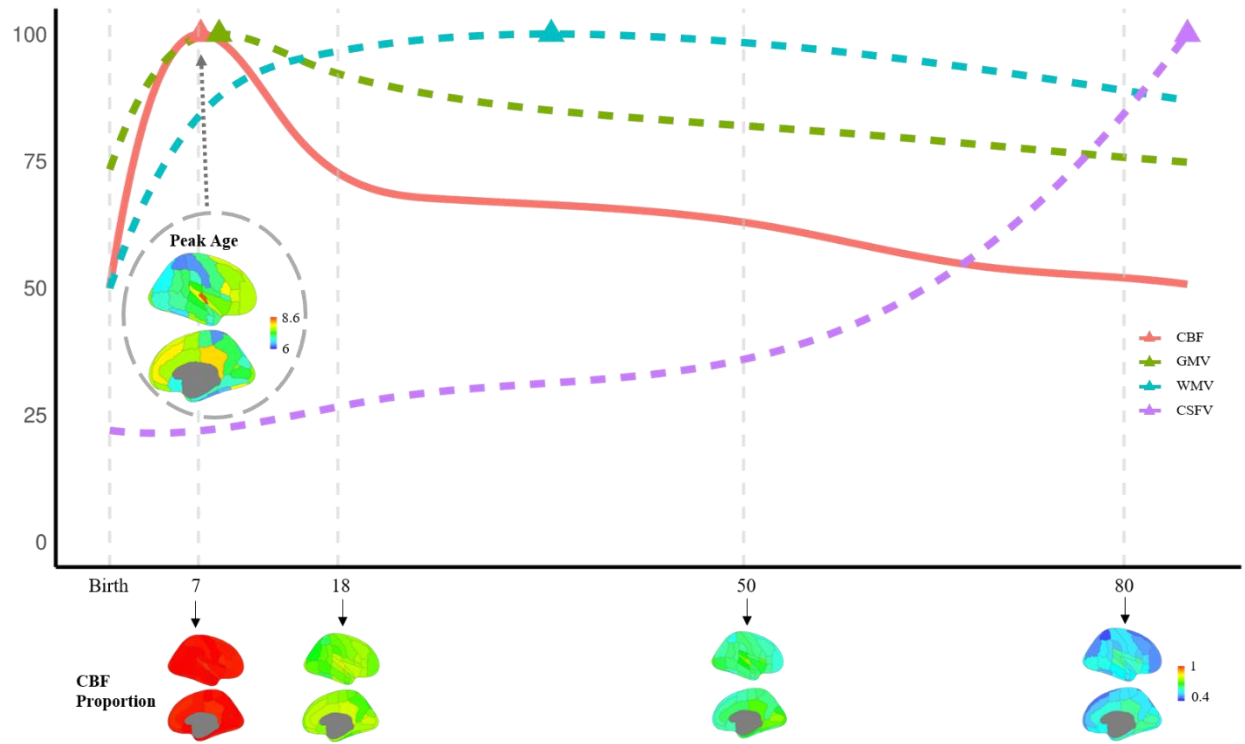

Figure 2. **Neurodevelopmental milestones.** A graphical summary of the normative trajectories of the median (50th percentile) for cerebral perfusion, gray matter volume, white matter volume, and cerebrospinal fluid volume across the lifespan. Triangles indicate the peak age for each phenotype, defined as the maximum of the median trajectories. The peak ages for regional cerebral perfusion are color-coded, with warmer colors indicating regions that mature later. Below the graph, regional perfusion maps illustrate the relative retention of cerebral perfusion at different life stages (e.g., early childhood, adolescence, adulthood, and old age), with warmer colors representing regions retaining a higher proportion of their peak perfusion.

**Characterizing individual CBF heterogeneity in brain disorders using cerebral perfusion normative models**

We assess the clinical utility of normative CBF modeling, we examined individual deviation scores derived from CBF data across four neurological and psychiatric disorders —AD, MCI, FTD, and MDD (Figure 3). These deviation scores quantified the degree of atypicality in cerebral perfusion patterns, revealing disorder-specific deviations.

At the global cerebral perfusion level (Figure 3, Panel A-B), a significant portion of patients demonstrated extreme negative deviations in cortical perfusion, with 14.1% of AD (Z = -1.57 ± 0.745), 13.6% of FTD (Z = -1.65 ± 0.734), 9.7% of MCI (Z = -1.09 ± 0.909), and 6.5% of MDD (Z = -1.15 ± 0.811) cases falling below normative thresholds. A similar pattern was observed in white matter perfusion, where 3.5% of AD (Z = -1.05 ± 0.856), 2.3% of FTD (Z = -0.93 ± 0.738), 5.2% of MCI (Z = -0.80 ± 0.936), and 3.8% of MDD patients (Z = -0.826 ± 0.818) exhibiting extreme deviations. Subcortical perfusion showed notable deficits in 10.6% of AD (Z = -1.43 ± 0.732), 9.1% of FTD (Z = -1.445 ± 0.783), 10.8% of MCI (Z = -1.04 ± 0.914), and 5.3% of MDD patients (Z = -1.12 ± 0.832).

At the network level (Figure 3, Panel A-B), disorder-specific patterns of perfusion deficits emerged. AD patients displayed significant perfusion decreases in the dorsal attention network (DAN, Z = -1.69 ± 0.771, 25.6% extreme negative deviation), frontoparietal network (FPN, Z = -1.74 ± 0.766, 24.7%), and default mode network (DMN, Z = -1.70 ± 0.793, 25%). FTD patients exhibited even greater deficits in: ventral attention network (VAN) (Z = -1.62 ± 0.844, 20.5%), FPN (Z = -1.83 ± 0.804, 25%), and DMN (Z = -1.85 ± 0.837, 34.1%). MCI patients demonstrated more heterogeneous deficits across multiple networks, with higher variability in network-specific Z-scores. In contrast, fewer than 10% of MDD patients showed extreme perfusion decreases across networks, suggesting relative stability in perfusion patterns within this group. Regional differences (Figure 3, Panel C-D) are detailed in Supplementary Tables 4.

These findings highlight the heterogeneity in cerebral perfusion profiles across brain disorders, underscoring the utility of normative models in capturing individual deviations from typical perfusion patterns. These deviations provide insights into disease-specific pathophysiology and hold potential for clinical applications, including subtype identification, classification, and prediction of clinical outcomes.

Using individual deviation scores, we identified two MCI subtypes and explored their predictive value for disease progression using cognitive score and diagnostic information (Figure 3, Panel E): MCI-N (MCI normal cerebral perfusion) and MCI-A (MCI AD like cerebral perfusion). Baseline (within 90 days to first ASL scan) Mini-mental state examination (MMSE) scores (Estimate = 0.766, t = 1.768, p = 0.078) and clinical dementia rating (CDR) scores (Estimate = -0.47625, t = 1.631, p = 0.103) did not differ significantly between the two clusters. However, individuals in the MCI-A group exhibited a steeper decline in MMSE scores (Estimate = -0.335, t = 4.156, $p < 0.001$) and a steeper increase in CDR scores (Estimate = 0.175, t = -3.133, p = 0.0018) over time, indicating faster cognitive decline. Logistic regression further revealed that individuals in the MCI-N group had significantly lower odds of converting to AD compared to the MCI-A group (OR = 0.45, 95% CI = [0.22, 0.94], p = 0.031), highlighting the clinical relevance of identifying these subtypes.

We assessed the classification performance of each disorder based on the area under the curve (AUC) derived from individual deviation scores after 1,000 permutation tests (Figure 3, Panel F). The classification model demonstrated high accuracy for FTD (mean AUC = 87.67, $p < 0.001$) and AD (mean AUC = 80.24, $p < 0.001$). For MCI, the classification performance was significant but lower in accuracy (mean AUC = 70.90, p = 0.001). In contrast, the model's performance for MDD was not significant (mean AUC = 54.88, p = 0.272), likely due to the imbalanced sample size between MDD and healthy controls.

In addition to classification, we examined the predictive value of perfusion deviations for clinical outcomes (Figure 3, Panel G). Perfusion deviations significantly predicted CDR scores in AD ($R^2 = 0.117$, p = 0.022) and MCI ($R^2 = 0.046$, p = 0.004). However, no significant association was found in FTD ($R^2 = 0.003$, $p = 0.420$). In MDD, perfusion deviations demonstrated a modest but significant association with Hamilton Depression Rating Scale (HDRS) scores ($R^2 = 0.012$, $p = 0.045$).

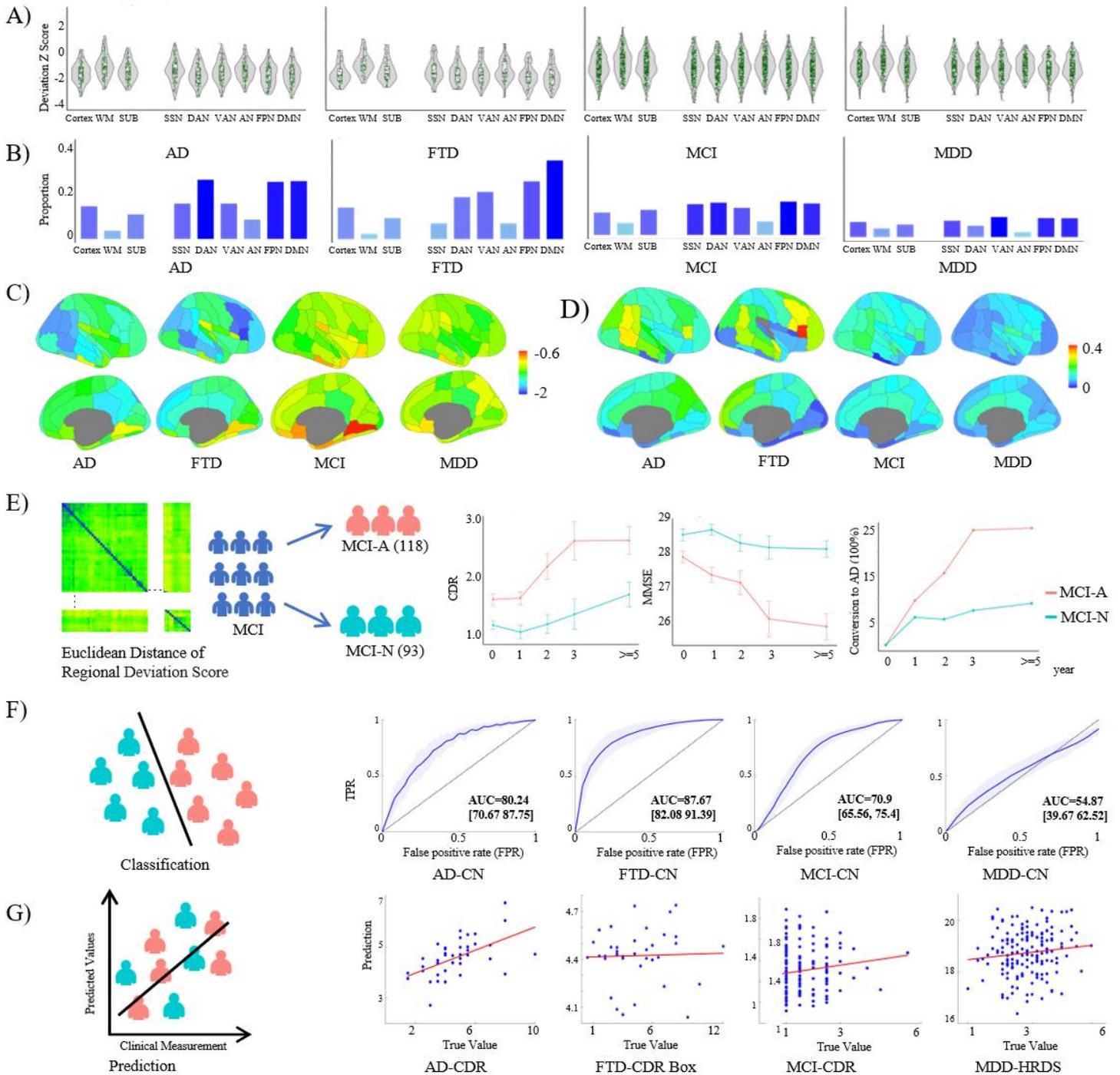

**Figure 3. Clinical Relevance of Perfusion Deviation Patterns in Brain Disorders.**
A) Violin plots illustrating deviation scores across global and network levels for AD,

FTD, MCI, and MDD. B) Proportion of extreme deviations (z < -2.3) at global and network levels across the four disorders. C) Regional deviation score maps depicting spatial patterns of perfusion deviations in AD, FTD, MCI, and MDD. D) Regional maps highlighting extreme negative deviations (Z < -2.3) across the four disorders. E) Identification of MCI subtypes (MCI-A and MCI-N) using Euclidean distances of regional deviation scores. Subtypes show distinct clinical trajectories for CDR, MMSE, and conversion rates to AD over time. F) Classification performance of brain disorders (AD, FTD, MCI, and MDD) compared to controls using receiver operating characteristic (ROC) curves based on deviation scores. G) Predictive value of perfusion deviation scores for clinical outcomes, including CDR in AD, CDR Box in FTD, CDR in MCI, and HDRS in MDD.

**Longitudinal Cerebral Perfusion in brain disorders**

We constructed new normative modeling of the longitudinal cerebral perfusion changes in neurodegenerative diseases, including AD, MCI, and CN (Figure 4). Changes in deviation scores over time provided valuable insights into the typicality or atypicality of aging-related perfusion patterns, revealing disorder-specific progression trajectories (Figure 4A-B).

At baseline, the total extreme negative proportion (TNP) of cerebral perfusion deviations (Z < -2.3) was significantly higher in AD (Estimate = 4.234, t = 2.773, p = 0.006) and MCI (Estimate = 2.053, t = 2.072, p = 0.039) compared to CN. Over time, TNP increased significantly in AD compared to CN (Estimate = 2.752, t = 2.260, p = 0.009), indicating a faster progression of perfusion abnormalities in AD. However, the rate of TNP increase was not significantly different between MCI and CN. Regionally, MCI and AD exhibited distinct progression trajectories in perfusion deviation scores (Figure 4C, left and right), and detailed statistical differences between the groups are presented in Supporting Table S5.

We further examined longitudinal changes in TNP between stable MCI and progressive MCI. While no significant difference in TNP was observed at baseline (p = 0.47), progressive MCI exhibited a significantly faster increase in TNP over time compared to stable MCI (Estimate = 1.629, t = 3.643, p = 0.0003), highlighting distinct trajectories within the MCI group (Figure 4D, left). Progressive and stable MCI subgroups also demonstrated different regional patterns in deviation score trajectories over time (Figure 4D, right). Statistical differences between the groups at the regional level are presented in Supporting Table S6.

Finally, we analyzed progression trajectories in MCI-A (AD like perfusion changes) and MCI-N (perfusion similar to NC) subtypes. Baseline TNP was significantly higher in MCI-A compared to MCI-N (Estimate = 6.41, t = 4.527, p < 0.001). However, the rate of TNP increase over time was not significantly different between the subtypes (p = 0.527). Both subtypes showed significant increases in TNP over time (MCI-A: $\beta$ = 0.611 [0.293, 0.929], p = 0.0002; MCI-N: $\beta$ = 0.773 [-0.0125, 1.56], p = 0.05; Figure 4E, left). Distinct regional progression patterns in deviation scores were also observed between the subtypes (Figure 4E, right). Regional progression statistics are detailed in Supporting Table S7.

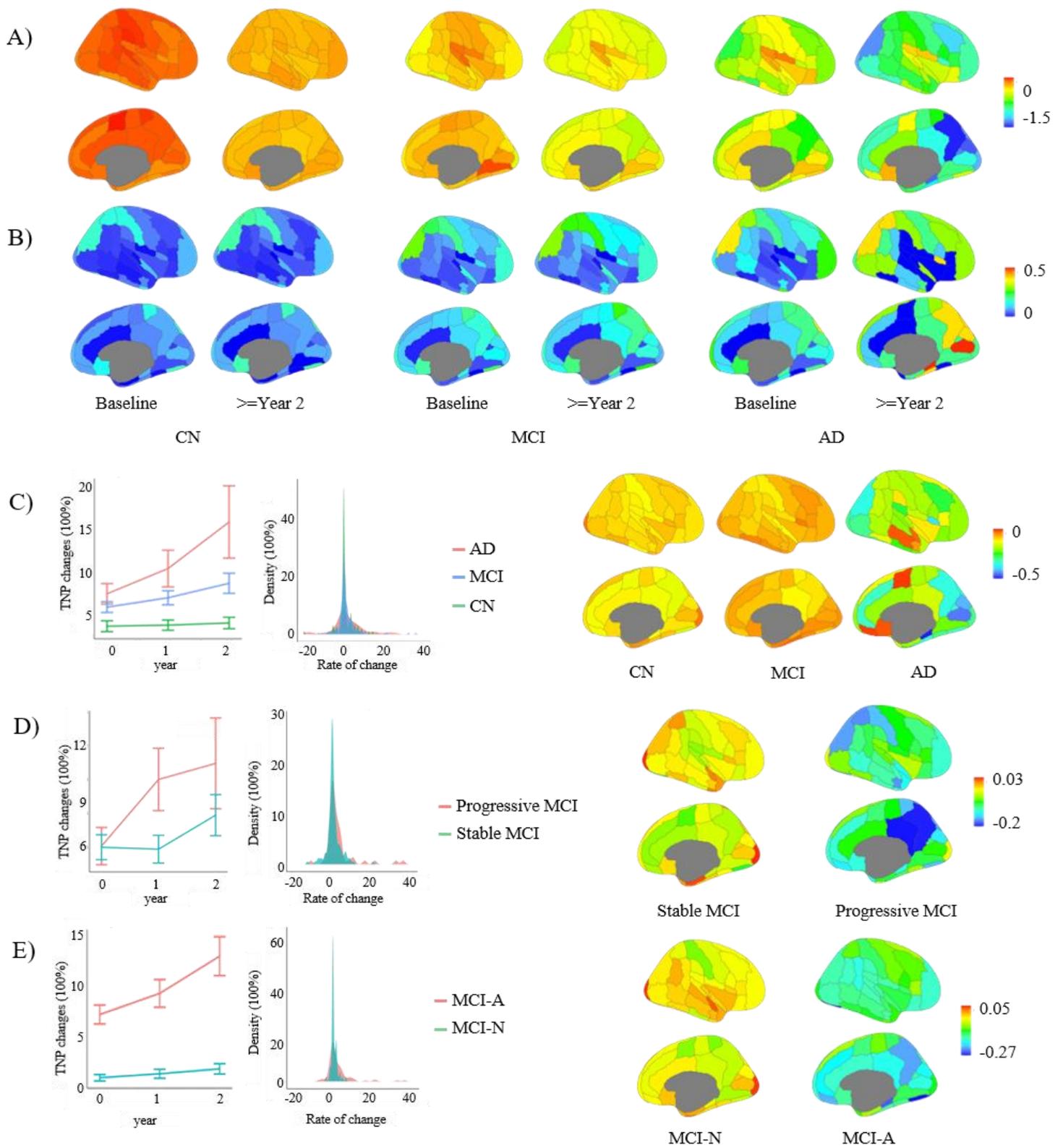

**Figure 4. Longitudinal perfusion progression in Brain Disorders.** Panel A, longitudinal perfusion deviation score maps at the regional level in CN, MCI, and AD

groups at baseline and follow-up. Panel B, longitudinal extreme negative perfusion deviation maps (Z < -2.3) at the regional level in CN, MCI, and AD groups at baseline and follow-up. Panel C: Perfusion change rates in CN, MCI, and AD groups. The left panel shows changes in total extreme negative proportion (TNP) over two years, while the right panel illustrates regional variability in perfusion change rates. Panel D: Perfusion change rates in progressive vs. stable MCI. The left panel shows changes in TNP between the two subgroups, while the right panel displays regional differences in perfusion change rates. Panel E: Perfusion change rates in MCI-A vs. MCI-N subtypes. The left panel shows changes in TNP over time for the two subtypes, while the right panel illustrates regional variability in perfusion change rates.

**Sensitivity analyses**

The lifespan growth patterns of cerebral perfusion were validated at the global, network, and regional levels using multiple rigorous analysis strategies. Each validation approach demonstrated strong alignment with the main results, confirming the robustness of the findings:

1. **Data Quality Validation**: Analyses were repeated with a stricter quality control threshold (QEI > 0.15) to assess the impact of data quality.
2. **Balanced Sampling**: A balanced sampling strategy was employed to address potential biases from uneven sample and site distributions across ages. Participants and site numbers were matched uniformly by resampling 1,000 times.
3. **Reproducibility Testing**: A split-half approach was used to validate the reproducibility of the results by dividing the dataset into two equal halves and comparing the outcomes.
4. **Bootstrap Resampling**: A bootstrap resampling analysis (1,000 iterations) was conducted to examine the influence of data sample variability.
5. **Site-Specific Effects**: A leave-one-site-out (LOSO) analysis was performed to evaluate the potential impact of specific sites on the observed patterns.

The results of these validation strategies were quantitatively compared to the primary findings and demonstrated consistent growth patterns across all analyses. Detailed results are provided in Supplementary Tables 8–13.

**Discussion**

This study is the first to comprehensively characterize cerebral perfusion growth patterns across the lifespan (birth to 85 years) using large multisite datasets. By integrating structural MRI and ASL MRI data, we mapped nonlinear growth trajectories at global, network, and regional levels, revealing critical milestones of perfusion maturation and age-related decline. Our normative models establish typical cerebral perfusion ranges, enabling precise detection of individual deviations and offering insights into both healthy aging and disease-specific patterns. The longitudinal analysis captured dynamic perfusion changes over time, providing a framework for understanding disease progression and transitions between clinical stages. These findings underscore the value of cerebral perfusion-based normative models as essential tools for studying brain development, aging, and the pathophysiology of neurodegenerative and psychiatric disorders.

At the global level, our study confirmed a rapid postnatal increase in cerebral perfusion, followed by a gradual decline with age, with peak age identified through precise confidence intervals derived from bootstrap and balanced resampling analyses. This early surge in cerebral perfusion is essential for meeting the heightened metabolic demands of the developing brain, supporting fundamental processes such as synaptogenesis, dendritic arborization, myelination, and glial proliferation (A et al., 2021; Ouyang et al., 2024; Silbereis et al., 2016). During childhood, the brain accounts for up to 44% of the body's total energy consumption (A et al., 2021; Blüml et al., 2013) , far surpassing the adult metabolic demands. This elevated perfusion ensures the delivery of oxygen and nutrients necessary for dynamic plasticity underlying sensory, cognitive, and motor development (Hijman et al., 2024; Ouyang et al., 2024). At the network level, the maturation of the DMN is facilitated by increased perfusion in DMN regions, reflecting the growing connectivity and functional specialization of this network (Yu et al., 2023). The structural development of the cerebral vasculature, including increased intracranial arterial diameters observed early in life, further supports these metabolic needs (Taylor et al., 2022). Disruptions in global and regional cerebral perfusion have been associated with neurodevelopmental delays and multiple infant disorders, highlighting the importance of balanced perfusion for typical brain development (Hijman et al., 2024; Mahdi et al., 2018; Nagaraj et al., 2015).

Following the rapid increase in cerebral perfusion during infancy, perfusion peaks around the age of 7 years and gradually declines throughout the life. This trajectory aligns with previous findings indicating the peak GMV occurs in early childhood, closely aligning with peak cerebral perfusion (Bethlehem et al., 2022). Sustaining brain volume during this period is highly energy-dependent, as daily glucose utilization by the brain also reaches its highest levels around this age (Kuzawa et al., 2014). Notably, brain glucose demand is inversely related to body growth from infancy to puberty, indicating that the substantial metabolic costs of neurodevelopment may be partially at the cost of a slowing of somatic growth (Aronoff et al., 2022; Kuzawa et al., 2014; Vandekar et al., 2017). During adolescence, as GMV decreases and WMV expands, the associated reduction in cerebral perfusion reflects a transition toward greater neural efficiency (Satterthwaite et al., 2014). This shift is reflected by the increasing neurovascular coupling during childhood, indicating improved coordination between neural activity and vascular responses (Schmithorst et al., 2015). After the phase of synaptic overproduction, synaptic pruning selectively eliminates redundant synapses to achieve optimal network efficiency (Nikiruy et al., 2024; Tierney & Nelson, 2009). Notably, this process follows a regionally specific timeline, with auditory regions completing pruning by around 6 years of age, whereas higher-order cognitive areas, such as the frontal cortex, continue pruning into adolescence (Dow-Edwards et al., 2019; Tierney & Nelson, 2009). These spatially inhomogeneous development patterns are consistent with our findings of a later cerebral perfusion peak age in frontal regions, reflecting the protracted development of higher-order cognitive and executive functions.

Significant sex differences in cerebral perfusion have been consistently reported across multiple studies (Alisch et al., 2021; Liu et al., 2016; Satterthwaite et al., 2014; Vandekar et al., 2017) , highlighting the critical influence of biological, hormonal, structural, and cardiovascular factors on the brain's vascular and metabolic processes. Anatomically, sex-based variations in brain size, gray matter volume, and cortical thickness contribute to distinct perfusion patterns (Bethlehem et al., 2022; Dukart et al., 2018). Hormonal influences, particularly estrogen and progesterone, play a central role in modulating cerebral blood flow, with estrogen's vasodilatory effects contributing to the higher perfusion levels typically observed in females (Cote et al., 2021; Krause et al., 2006; Liu et al., 2016). Cardiovascular factors further shape these differences, as sex-related variations in systemic blood pressure, heart rate, and vascular compliance influence cerebral perfusion regulation (Barnes & Charkoudian, 2021).

Cerebral perfusion normative modeling provides a powerful tool for capturing individual differences, mapping disease-specific patterns, and characterizing variability in disorders such as AD, FTD, MCI, and MDD. While previous studies have extensively investigated group-level cerebral perfusion differences between these disorders and healthy controls (Alsop et al., 2000; Hu et al., 2010a; Liao et al., 2017; van Dinther et al., 2024; Wang et al., 2013; Wang et al., 2007; Wei et al., 2018; Ze Wang, 2017) , this study is the first to employ a normative modeling framework, offering several key advantages and novel insights. First, normative modelling enables individualized inference, revealing disease heterogeneity beyond group averages. By leveraging large multisite datasets, this method provides more precise estimates of individual deviations from the median. For instance, in FTD, although the mean deviation z-score was similar between the FPN and the DMN, the proportion of extreme negative deviations was significantly higher in the DMN (34%) compared to the FPN (25%), suggesting greater consistency of alterations in the FPN but higher heterogeneity in the DMN. Second,

age- and sex-normalized cerebral perfusion facilitates transdiagnostic comparisons and disease-specific analyses. Segal et al. Segal et al. (2023) previously demonstrated the utility of such approaches in identifying shared and distinct brain abnormalities across psychiatric disorders. In our study, we observed distinct patterns of cerebral perfusion alterations across neurodegenerative diseases and psychiatric conditions. Neurodegenerative diseases like AD, FTD, and MCI exhibited greater degrees of perfusion alterations, characterized by higher proportions of extreme deviations and lower mean z scores compared to MDD. These findings are consistent with evidence suggesting that neurodegenerative diseases are more prominently associated with neurovascular dysfunction (Solis et al., 2020). Third, the normative model enhances sensitivity to subtle perfusion changes, increasing statistical power and improving discrimination between disease-related and non-disease-related variability (Rutherford et al., 2022). This advantage is particularly relevant for early diagnosis, as demonstrated in our MCI analysis. We identified two subgroups: MCI-A, characterized by decreased cerebral perfusion similar to AD, and MCI-N, with perfusion patterns resembling those of normal controls. While baseline cognitive impairment did not differ significantly between the subgroups, their trajectories diverged substantially over time. MCI-A exhibited greater declines in cognition and cerebral perfusion and a higher likelihood of conversion to AD, highlighting the predictive value of cerebral perfusion deviations for disease progression (Duan et al., 2021; Solis et al., 2020; Wolters et al., 2017). Finally, we demonstrated the practical applications of deviation scores in disease classification and prediction. These findings underscore the importance of individualized cerebral perfusion assessments for understanding disease heterogeneity, predicting progression, and guiding early interventions.

The longitudinal analysis of cerebral perfusion using normative modeling offers critical insights into the progression trajectories of neurodegenerative diseases, particularly in AD and MCI, compared to CN individuals. Over time, the significantly faster decline in cerebral perfusion and increase in TNP (total negative perfusion) in AD relative to CN highlights the progressive nature of perfusion abnormalities in AD. This pattern reflects accelerated neurovascular and metabolic dysfunction, consistent with previous findings of increasing brain atrophy in AD using normative models (Verdi et al., 2024). Given that accelerated disease progression is a hallmark of AD (Leung et al., 2013), these findings underscore the value of normative modeling for early diagnosis and intervention. In contrast, MCI exhibited notable heterogeneity, with distinct trajectories observed in stable and progressive MCI subgroups. Progressive MCI showed a faster increase in TNP over time, while the absence of baseline differences between stable and progressive MCI suggests that longitudinal modeling can uncover subtle progression patterns that may not be detectable in cross-sectional analyses. Furthermore, the longitudinal trajectories of cerebral perfusion in stable and progressive MCI, as well as in the MCI-A and MCI-N subgroups, suggest bidirectional associations between cerebral perfusion decline and the likelihood of AD conversion. This aligns with previous studies linking increased brain atrophy to a heightened risk of AD conversion (Verdi et al., 2023; Verdi et al., 2024). Collectively, these findings demonstrate the utility of integrating cross-sectional and longitudinal normative modeling to capture disease progression dynamics, inter-individual heterogeneity, and regional-specific patterns in neurodegeneration.

Several challenges and implications should be considered in the context of this study. First, the estimation of CBF can be influenced by various ASL sequence factors, such as the use of PASL versus PCASL, single versus multiple PLDs, the inclusion of a separate M0 image or one generated from control images, and the application of background suppression (Luis Hernandez-Garcia, 2022; Ze Wang, 2022). Although we accounted for sequence effects by including them as fixed factors in the normative modeling, this approach enhances generalizability but does not fully mitigate the inherent limitations of the primary study designs. Developing a more standardized and quantitative pipeline for cerebral perfusion assessment across multicenter studies is needed to ensure robust and comparable results. Second, similar to prior normative modeling studies using structural MRI (Bethlehem et al., 2022), the current neuroimaging datasets disproportionately represent populations from Europe and North America. Future research should prioritize the inclusion of more diverse neuroimaging cohorts, ensuring balanced representation across ethnic groups and reducing geographic bias. Third, ASL data included in this study were acquired with a low resolution. Some were with limited brain coverage to include the full visual cortex and cerebellum. As more and more research sites are using the state-of-arts ASL MRI sequence developed by us (Chang et al., 2017; Vidorreta et al., 2014; Vidorreta et al., 2017; Vidorreta et al., 2012) and others (Alsop et al., 2014), high-resolution ASL CBF maps may become more available, allowing us to explore finer-grained cerebral perfusion atlases. Fourth, the age distribution of participants in the datasets was uneven, with certain age ranges, such as 0–5 years and 30–40 years, being underrepresented. Including data from fetal and neonatal stages would provide a more complete understanding of cerebral perfusion across the lifespan (Wang, Fernandez-Seara, et al., 2008). Fifth, this study focused exclusively on CBF though ASL MRI can provide several other neurovascular measures in addition to CBF, including arterial transit time, blood-brain-barrier water exchange time or rate, and vascular health (Li et al., 2022; Li & Wang, 2023). These parameters, however, depend on long ASL MRI scans with different PLDs and TEs or pre-conditions such as diffusion-weighted vasculature water saturations, which were not-available in most of datasets included in this paper. When data will be available, expanding future models to incorporate these metrics could yield a more comprehensive understanding of brain physiology. Sixth, while we demonstrated the application of regional deviation scores in subgroup identification, classification, and prediction, the reliability and utility of these applications could be greatly enhanced with advancements in artificial intelligence algorithms and higher-quality cerebral perfusion data. These developments may ultimately facilitate the translation of these tools into clinical practice. Finally, the cerebral perfusion-based growth charts established in this study are intended to serve as a dynamic resource. As more high-quality datasets on brain metabolism become available, these lifespan normative growth models can be refined and updated, further enhancing their utility for research and clinical applications.

**Conclusion**

This study establishes normative cerebral perfusion charts across the lifespan, providing a robust framework to quantify individual variability and detect deviations linked to neurodevelopmental and neurodegenerative processes. By integrating cross-sectional and longitudinal analyses, these models offer insights into brain development, aging, and disease progression, enabling precise characterization of heterogeneity and prediction of clinical trajectories. These findings highlight the potential of perfusion-based models as valuable tools for research and clinical applications.

**Method**

**Datasets and Participants**

To delineate the normative growth of cerebral perfusion in the human brain, we aggregated multisite neuroimaging datasets, each containing both 3T structural MRI and resting-state ASL data. For participants with multiple test-retest scans or longitudinal data, only the session with the highest quality score was included. Written informed consent was obtained from participants or their legal guardians, and all recruitment procedures were approved by the local ethics committees overseeing each dataset.

**Image quality control process**

To ensure the reliability of imaging data, all T1-weighted structural MRI and ASL scans exhibiting severe artifacts or partial brain coverage were independently reviewed and excluded by X.Z. and Y.L. Additionally, data quality was further assessed using the cerebral blood flow quality evaluation index (QEI) (Dolui et al., 2024) , a metric ranging from 0 to 1, with higher values indicating better quality. The QEI captures three critical aspects of valid CBF maps: 1) **Structural similarity**: The degree of correlation between brain structure and CBF, as structure and function are typically aligned; 2)

**Spatial variability**: The variability of CBF values across voxels within each tissue type. High variability suggests data noise or insufficient post-labeling delay. All the participants with QEI less than 0.1 were excluded for further analysis and we used several thresholds for validation.

**Image preprocessing**

**Structural data preprocessing**

Publicly available, containerized HCP structural preprocessing pipeline was used (Glasser et al., 2013). Briefly, this pipeline consists of two stages: (1) The PreFreeSurfer stage: This stage normalizes anatomical MRI data and includes preprocessing steps such as brain extraction, denoising, and bias field correction of T1-weighted images. (2) The FreeSurfer stage: This stage generates cortical surfaces from the normalized anatomical data. Key steps include anatomical segmentation, construction of pial, white, and mid-thickness surfaces, and surface registration to the standard atlas. The preprocessed T1 data was used for further ASL CBF estimating, registration and providing brain volume information.

**ASL data preprocessing**

ASLtbx (Li et al., 2018; Wang, 2012; Wang, Aguirre, et al., 2008) was used to process the ASL MRI data using standard processing pipelines. The ASL label and control images were motion-corrected, temporal confound filtered, and spatially smoothed. Quantitative CBF maps were calculated in physiological units (mL/100 g/min) using the general kinetic model:

$$CBF = \frac{\Delta M}{2 * M0 * \lambda * \alpha * \beta * e^{-\sigma/T}} \quad (1)$$

, where $\Delta M$ is the difference between control and label images, $M0$ is the equilibrium magnetization of arterial blood, $\lambda$ is the brain-blood partition coefficient, $\alpha$ is the labeling efficiency, $\beta$ is the bolus duration, $\sigma$ is the post-labeling delay (PLD), and $T$ is the longitudinal relaxation time of blood (1.65 s at 3T). M0 images were explicitly acquired in the 3D background suppressed ASL data, and the pulsed ASL (PASL) data.

For the 2D background unsuppressed pseudo continuous Arterial Spin Labeling (pCASL) data, the mean of the control images from the ASL time series was used as the M0 image. Partial volume correction was performed using tissue segmentation results based on the structural T1-weighted images.

Functional-to-standard image registration was performed to align processed CBF maps to MNI152 space for region-specific analyses. First, the CBF map was linear registered to the subject's T1-weighted structural image. The T1-weighted image was then nonlinear aligned to the MNI152 standard space. The CBF image was subsequently transformed to standard space by combining the functional-to-structural affine matrix and the nonlinear warp. All registration outputs were visually inspected to ensure accuracy. Regional CBF values were measured using the Harvard-Oxford brain atlas, encompassing 48 cortical and 8 subcortical regions (Desikan et al., 2006). Additionally, network-level CBF values were explored using Yeo's atlas, focusing on networks such as the DMN, FPN, affective network (AN), VAN, DAN, and somatosensory network (SSN) (Yeo et al., 2011). Due to potential sequence issues, the visual network was excluded from the analysis.

**Modeling normative growth curves across the lifespan**

To estimate normative growth patterns for cerebral perfusion metrics in healthy individuals across cohorts, we applied GAMLSS to the cross-sectional data using the **gamlss** package (version 5.0-6) in R 4.2.0 (Borghi et al., 2006; Stasinopoulos & Rigby, 2008). The GAMLSS procedure consisted of two steps: identifying the optimal data distribution and determining the best-fitting model parameters for each global cerebral perfusion metric (Sun et al., 2024). Using these metric-specific GAMLSS models, we obtained nonlinear normative growth curves and their first derivatives. Furthermore, the sex-stratified growth patterns were assessed. Nonlinear normative growth curves and their first derivatives were generated, with sex-stratified growth patterns also examined. The robustness of these growth curves was validated using bootstrap resampling, leave-one-study-out analysis, balanced resampling, and split-half replication (Bishop, 2006).

**(i). Model data distributions**

We assessed 28 continuous distribution families for their fit to the data using global cortical CBF as the reference metric. Model fits were evaluated using the Bayesian Information Criterion (BIC) (Neath & Cavanaugh, 2012), where lower BIC values indicate superior model performance. Among the evaluated distributions, the Generalized Beta 2 (GB2) distribution consistently provided the best fit across all models (see Supplementary Figure).

**(ii). The GAMLSS framework.**

The GAMLSS framework was applied with CBF values as the dependent variable, age as a smoothing term (using B-spline basis functions), and sex as a fixed effect. To account for variability introduced by ASL sequences, the ASL sequence type was also included as a fixed effect, while dataset site was treated as a random effect (Sun et al., 2024). The GB2 distribution, which has four parameters: median ($\mu$), coefficient of variation ($\sigma$), skewness ($v$), and kurtosis ($\tau$), was chosen to fit the data distribution. Each CBF values, denoted by $y$, was modeled as:

$$y = GB2\ (\mu, \sigma, v, \tau) \qquad (2)$$

$$\mu = f_\mu\ (age) + \beta_\mu^1\ (sex) + \beta_\mu^2\ (sequence) + Z_\mu\ (site) \qquad (3)$$

$$\sigma = f_\sigma\ (age) + \beta_\sigma\ (sex) \qquad (4)$$

$$v = \beta_v \qquad (5)$$

$$\tau = \beta_\tau \qquad (6)$$

To explore the age-related trends in global and network-level CBF values, three GAMLSS models with different degrees of freedom (df = 5–9) were tested for the B-spline basis functions in the μ(location) and σ (scale) parameters. The optimal model was selected based on the lowest BIC value. Across all analyses, a consistent df = 8 was identified as optimal. Regional CBF models also used df = 8 to maintain consistency in identifying potential peak ages. Following prior studies, only intercept terms were included for the ν and τ parameters (Di Biase et al., 2023). For model estimation, we used a default convergence criterion of log-likelihood < 0.001 between iterations, with a maximum of 500 iteration cycles.

**(iii). Sex differences across the lifespan.**

To assess the influence of sex on CBF across the lifespan, we included sex as a fixed effect in the GAMLSS model. Adjusted mean (μ) and variance (σ) coefficients, along with their standard errors, T-values, and P-values, were computed for the sex variable (see Supplementary Tables 3 and 4 for global and network-level CBF, respectively). The T-value, calculated as the coefficient divided by its standard error, was used to test the null hypothesis that the sex has no significant effect on CBF. To ensure robust estimates of sex-specific effects on CBF growth trajectories, we adjusted for key covariates, including age, ASL sequence type, and scanner site as a random effect.

**Sensitivity analysis of normative models**

To validate the lifespan normative growth patterns at global, network, and regional levels, we conducted multiple sensitivity analyses. These analyses addressed critical methodological concerns, including data quality, sample size imbalance, site distributions variability, model replication, model stability, and the influence of specific sites. By systematically testing these factors, we ensured the robustness and generalizability of our normative models across diverse datasets and methodological conditions.

**(i). CBF data quality assurance using QEI > 0.15.**

To ensure the robustness of our findings against variations in ASL data quality, a stricter quality control threshold was applied, excluding participants with a QEI below 0.15. This stricter threshold was selected based on previous research emphasizing the critical role of ASL data quality in brain imaging studies (Li & Wang, 2023). Normative model analyses were then re-run to verify consistency with the original results, ensuring the reliability of our findings.

**(ii). Balanced resampling analysis.**

To mitigate potential biases from uneven sample sizes and site distributions across age groups, a balanced sampling strategy was employed. The lifespan was segmented into 17 five-year age bins spanning birth to 85 years, with the smallest group containing 64 participants (ages 35–40). To achieve balanced representation across age groups, random sampling was used to equalize all age group to 64 participants. This procedure was repeated at ratios of 1×, 1.5×, and 2× the size of the smallest group size (64 here) to assess the robustness of the findings.

For global and network CBF values, sampling was repeated 1,000 times on a pool of 9,800 participants. Each iteration included a random subset of 1,300 participants, and GAMLSS models were refitted to generate 1,000 growth curves per metric. Confidence intervals were calculated for the growth curves, the peak of the 50th centile, and the correlations between the resampled median centile curves and the original cohort's median centile curve.

**(iii). Split-half replication analysis.**

To evaluate model reproducibility and performance, we employed a split-half validation. The dataset was randomly divided into two equal halves, stratified by site. One half was used to train the GAMLSS model, and the other half was used to evaluate the model's goodness of fit. This process was repeated with the roles of training and testing sets reversed.

Model fit was assessed using R-squared ($R^2$) for central tendency and quantile randomized residuals (randomized z-scores) for calibration. The Shapiro-Wilk test evaluated the normality of residuals, with $W$ values close to 1 indicating good fit. Higher-order moments, including skewness (values near 0 indicate symmetry) and kurtosis (values near 0 indicate light tails), provided further insights into model fit. This procedure was repeated 1,000 times, generating median and 95% CI values for $R^2$, $W$, skewness, and kurtosis.

**(iv). Bootstrap resampling analysis.**

To assess the robustness of lifespan growth curves, we conducted 1,000 bootstrap repetitions with replacement sampling. The sampling preserved the age and sex proportionality of the original cohort by stratifying the lifespan into nine intervals. For each functional metric, 1,000 growth curves were generated, and 95% CIs were computed for the median (50th) centile curve and inflection points. The CIs were derived from the mean and standard deviation of the bootstrap growth curves and growth rates.

**(v). Leave-one-study-out (LOSO) analysis.**

To evaluate the influence of individual sites on the growth curves, LOSO analysis was performed. Samples from one site were excluded in each iteration, and the GAMLSS models were refitted. Growth curves and rates were then re-estimated, with mean and standard deviation values used to compute 95% CIs for both the growth curves and rates. The narrow CIs indicated that the models remained robust and consistent, even when data from individual sites were removed.

**Clinical relevance of CBF-based normative models in brain disorders**

To evaluate the clinical utility of lifespan perfusion models, this study analyzed quality-controlled structural MRI and ASL data from individuals with four brain disorders: MDD, MCI, AD, and FTD. Data were drawn from the EMBARC (MDD), ALLFTD (FTD; 5 sites), ADNI (AD and MCI), and UMB (MCI) datasets, including 28 CNs and 186 patients with MDD, 56 CNs and 44 patients with FTD, 213 CNs and 64 patients with AD, and 251 patients with MCI.

**(i). Individual deviation z scores**.

The standard protocol for normative modeling emphasizes the importance of including control samples from the same imaging sites as the patient data in the testing set to account for site effects. To ensure robust modeling, a stratified approach was adopted, incorporating site-specific CNs in the testing set. This approach enabled the identification and mitigation of site effects, minimizing confounding in case-control comparisons. All CNs from disease-specific datasets were randomly split into training and testing subsets, stratified by sex and site. Lifespan normative models of cerebral perfusion were constructed using the training set, which included half of the CNs and data from other large datasets.

The testing set, comprising the remaining CNs and all patient cases, served as an independent validation set for calculating deviation scores. Quantile scores relative to the normative model curves were calculated for each individual, followed by the computation of deviation z-scores. These z-scores were derived by transforming the fitted generalized beta distribution (GB2) quantiles into standard Gaussian z-scores using quantile randomized residuals. Extreme negative deviations were defined as $z < -2.3$, consistent with prior studies. Percentage maps of extreme deviations revealed substantial heterogeneity in deviation patterns among individuals within each disease group.

To ensure reliability, this process was repeated 100 times, generating 100 independent models and corresponding sets of deviation scores for both patients and testing CNs. The normality of the z-score distributions was evaluated using two-tailed Kolmogorov-Smirnov tests, which consistently yielded $p<0.05$ across all models and iterations. Subsequent analyses were based on these independently derived deviation scores for testing HCs and patient cohorts. The stability of individual deviation estimates was assessed by calculating pairwise Pearson correlation coefficients and mean squared errors (MSE) across the 100 independent models. Results demonstrated highly stable deviation scores within specific disease cohorts, with mean correlations exceeding 0.95 and mean MSE values below 0.2 for all metrics. Case-control comparisons and disease classification analyses were repeated across the 100 models to confirm robustness.

**(ii). Identification of MCI Subtypes Using individual perfusion deviations.**

Given the substantial individual heterogeneity in cerebral perfusion patterns and the limited number of patients, we employed a data-driven k-means clustering algorithm to identify subtypes of MCI. Deviation features for each patient included both regional and network-level CBF metrics. The Euclidean distance was used to calculate the similarity matrix across patients. The optimal number of clusters was determined to be between 2 and 8. To identify the final cluster count, we utilized the NbClust package, which computes 30 different clustering indices. The most frequently identified optimal cluster number was selected as the final solution. The clustering analysis revealed two distinct MCI subtypes. One subtype exhibited deviation patterns closer to those of AD, which we labeled MCI-A, while the other showed deviations more similar to CNs, labeled MCI-N. We then explored disease progression trajectories within these subtypes. A logistic regression analysis was conducted to assess the association between baseline subtype and the risk of conversion from MCI to AD during the follow-up period. The outcome variable was binary, indicating whether participants converted to AD (Converted = 1) or remained MCI (Converted = 0). Age and gender were included as covariates to account for potential confounding factors. The logistic regression model was fitted using a generalized linear model (GLM) with a binomial distribution (logit link function). The odds ratio (OR) and corresponding 95% CIs were calculated to quantify the effect of each predictor on conversion risk. Statistical significance was set at $p < 0.05$. To further evaluate the longitudinal cognitive changes in the identified subtypes, we applied a linear mixed model (LMM) to examine the interaction between MCI subtype and time on the MMSE scores and CDR scores. Age and gender were included as covariates in the model. The fixed effects included MCI subtype, time (defined as year), and their interaction to capture group differences in longitudinal changes. Time points were defined as whole-year intervals, with measurements within ±3 months assigned to the closest year (e.g., 11 months as Year 1, 26 months as Year 2). Different participants were as a random effect to account for repeated measures over time.

**(iii). Disease classification and prediction based on connectome-based deviations.**

We applied support vector machine (SVM; www.csie.ntu.edu.tw/~cjlin/libsvm/) analysis to evaluate the discriminative power of perfusion-based deviations in distinguishing patients from healthy controls (HCs). Model was repeated 1000 times using a 2-fold cross-validation framework, with training and testing sets alternated in each fold. Classification performance was assessed by plotting receiver operating characteristic (ROC) curves and calculating the area AUC. The statistical significance of AUC values was determined using a nonparametric permutation test (1,000 iterations), in which labels were randomly shuffled prior to implementing SVM and cross-validation. This procedure generated a null distribution of AUC values, and corresponding p-values were computed. Mean ROC curves and mean AUC values were derived by averaging results across the 1000 repetitions.

We employed support vector regression (SVR) with a linear kernel to predict clinical scores based on connectome-based deviations. A 2-fold cross-validation framework was used to estimate prediction accuracy. Each fold alternated as the training and test set. Similar to SVM, features were normalized in the training set, with the same parameters applied to the testing set. Predictive performance was evaluated using Pearson's correlation coefficients between predicted and observed clinical scores. Statistical significance was assessed via a nonparametric permutation test (1,000 iterations), with target scores shuffled before implementing SVR and cross-validation. This process yielded a null distribution of correlation coefficients, and p-values were computed. All analyses were conducted using the libsvm software.

**Longitudinal cerebral perfusion changes in brain disorders**

To investigate longitudinal changes in cerebral perfusion in brain disorders, this study utilized quality-controlled structural MRI and ASL data (visual inspection and QEI values) from individuals with at least one year of longitudinal neuroimaging follow-up. Prevent AD and Dalas lifespan were set as reference health control dataset, which includes 506 participants with 1430 scans, the ADNI were set as disease dataset, included 649 health controls with 213 scans, 256 MCI with 911 scans, 64 AD with 189 scans.

**(i). Deviation z scores at each scan.**

Deviation z-scores for each scan were calculated using a procedure similar to the previously described normative model. The reference healthy control dataset, combined with a randomly sampled 50% of healthy controls from the disease dataset, served as the training dataset. The remaining 50% of healthy controls and all patients were included in the testing dataset. This process was repeated 100 times, and the final deviation z-scores for each scan were averaged across the 100 iterations. The TNP was defined as the proportion of extreme negative deviations ($z < -2.3$) across all brain regions (58 cortical and 8 subcortical regions) (Verdi et al., 2024).

**(ii). Cerebral perfusion progression model**

The timepoint of each scan was recoded as the number of months after the first scan and categorized into yearly intervals (±3 months). LMMs were employed to examine the effect of timepoint on cerebral perfusion metrics. Dependent variables included TNP and z-scores for global, network, and regional perfusion, with timepoint as the independent variable. Group differences were assessed by modeling the interaction between group and timepoint effects. Age and gender were included as covariates, and individual participants were treated as random effects to account for repeated measures over time.

For MCI participants, progression models were analyzed based on diagnostic stability. Stable MCI was defined as participants who remained in the MCI diagnostic category throughout follow-up, while progressive MCI was defined as participants who converted to AD during the study period. Additionally, progression models were examined for the MCI-A and MCI-N subtypes to identify differences in longitudinal perfusion trajectories.

# Acknowledgements

Part of the data were from the ADNI, UK Biobank.

This work was supported by the National Institute on Aging [R01AG081693, R01AG070227, R21AG080518], and the National Institute on Biomedical Imaging and Bioengineering [R01 EB031080] and the University of Maryland Baltimore, Institute for Clinical & Translational Research (ICTR) [1UL1TR003098].

**Reference**

A, O., U, M., Lf, B., & A, G.-C. (2021). Energy metabolism in childhood neurodevelopmental disorders. *EBioMedicine*, *69*, 103474. https://doi.org/10.1016/j.ebiom.2021.103474

Alisch, J. S. R., Khattar, N., Kim, R. W., Cortina, L. E., Rejimon, A. C., Qian, W.,···Bouhrara, M. (2021). Sex and age-related differences in cerebral blood flow investigated using pseudo-continuous arterial spin labeling magnetic resonance imaging. *Aging (Albany NY)*, *13*(4), 4911-4925. https://doi.org/10.18632/aging.202673

Alsop, D. C., Casement, M., de Bazelaire, C., Fong, T., & Press, D. Z. (2008). Hippocampal hyperperfusion in Alzheimer's disease. *NeuroImage*, *42*(4), 1267-1274. https://doi.org/S1053-8119(08)00720-9 [pii]

10.1016/j.neuroimage.2008.06.006

Alsop, D. C., Dai, W., Grossman, M., & Detre, J. A. (2010). Arterial spin labeling blood flow MRI: its role in the early characterization of Alzheimer's disease. *Journal of Alzheimer's disease : JAD*, *20*(3), 871-880. https://doi.org/10.3233/JAD-2010-091699

Alsop, D. C., Detre, J. A., Golay, X., Gunther, M., Hendrikse, J., Hernandez-Garcia, L.,···Zaharchuk, G. (2014). Recommended implementation of arterial spin-labeled perfusion MRI for clinical applications: A consensus of the ISMRM perfusion study group and the European consortium for ASL in dementia. *Magnetic Resonance in Medicine*. https://doi.org/10.1002/mrm.25197

Alsop, D. C., Detre, J. A., & Grossman, M. (2000). Assessment of cerebral blood flow in Alzheimer's disease by spin-labeled magnetic resonance imaging. *Annals of neurology*, *47*(1), 93-100.

Aronoff, J. E., Ragin, A., Wu, C., Markl, M., Schnell, S., Shaibani, A.,···Kuzawa, C. W. (2022). Why do humans undergo an adiposity rebound? Exploring links with the energetic costs of brain development in childhood using MRI-based 4D measures of total cerebral blood flow. *International Journal of Obesity*, *46*(5), 1044-1050. https://doi.org/10.1038/s41366-022-01065-8

Barnes, J. N., & Charkoudian, N. (2021). Integrative cardiovascular control in women: Regulation

of blood pressure, body temperature, and cerebrovascular responsiveness. *Faseb j*, *35*(2), e21143. https://doi.org/10.1096/fj.202001387R

Bethlehem, R. A. I., Seidlitz, J., White, S. R., Vogel, J. W., Anderson, K. M., Adamson, C.,···Alexander-Bloch, A. F. (2022). Brain charts for the human lifespan. *Nature*, *604*(7906), 525-533. https://doi.org/10.1038/s41586-022-04554-y

Biagi, L., Abbruzzese, A., Bianchi, M. C., Alsop, D. C., Del Guerra, A., & Tosetti, M. (2007). Age dependence of cerebral perfusion assessed by magnetic resonance continuous arterial spin labeling. *J Magn Reson Imaging*, *25*(4), 696-702. https://doi.org/10.1002/jmri.20839

Bishop, C. (2006). *Pattern Recognition and Machine Learning*. Springer-Verlag.

Blüml, S., Wisnowski, J. L., Nelson, M. D., Jr., Paquette, L., Gilles, F. H., Kinney, H. C., & Panigrahy, A. (2013). Metabolic maturation of the human brain from birth through adolescence: insights from in vivo magnetic resonance spectroscopy. *Cereb Cortex*, *23*(12), 2944-2955. https://doi.org/10.1093/cercor/bhs283

Borghi, E., de Onis, M., Garza, C., Van den Broeck, J., Frongillo, E. A., Grummer-Strawn, L.,···Martines, J. C. (2006). Construction of the World Health Organization child growth standards: selection of methods for attained growth curves. *Stat Med*, *25*(2), 247-265. https://doi.org/10.1002/sim.2227

Camargo, A., Wang, Z., & Alzheimer's Disease Neuroimaging, I. (2021). Longitudinal Cerebral Blood Flow Changes in Normal Aging and the Alzheimer's Disease Continuum Identified by Arterial Spin Labeling MRI. *Journal of Alzheimer's disease*, *81*(4), 1727-1735. https://doi.org/10.3233/JAD-210116

Camargo, A., Wang, Z., & Initiative, A. s. D. N. (2023). Hypo-and hyper-perfusion in MCI and AD identified by different ASL MRI sequences. *Brain Imaging Behav*, 1-14.

Carsin-Vu, A., Corouge, I., Commowick, O., Bouzillé, G., Barillot, C., Ferré, J.-C., & Proisy, M. (2018). Measurement of pediatric regional cerebral blood flow from 6 months to 15 years of age in a clinical population. *Eur J Radiol*, *101*, 38-44. https://doi.org/10.1016/j.ejrad.2018.02.003

Chang, Y. V., Vidorreta, M., Wang, Z., & Detre, J. A. (2017). 3D-accelerated, stack-of-spirals acquisitions and reconstruction of arterial spin labeling MRI. *Magn Reson Med*, *78*(4), 1405-1419. https://doi.org/10.1002/mrm.26549

Chen, J. J., Rosas, H. D., & Salat, D. H. (2013). The Relationship between Cortical Blood Flow and Sub-Cortical White-Matter Health across the Adult Age Span. *PLoS One*, *8*(2), e56733. https://doi.org/10.1371/journal.pone.0056733

Chiappelli, J., Adhikari, B. M., Kvarta, M. D., Bruce, H. A., Goldwaser, E. L., Ma, Y.,···Hong, L. E. (2023). Depression, stress and regional cerebral blood flow. *J Cereb Blood Flow Metab*, *43*(5), 791-800. https://doi.org/10.1177/0271678x221148979

Churchill, N. W., Graham, S. J., & Schweizer, T. A. (2023). Perfusion Imaging of Traumatic Brain Injury. *Neuroimaging Clin N Am*, *33*(2), 315-324. https://doi.org/https://doi.org/10.1016/j.nic.2023.01.006

Clark, L. R., Berman, S. E., Rivera-Rivera, L. A., Hoscheidt, S. M., Darst, B. F., Engelman, C. D.,···Johnson, S. C. (2017). Macrovascular and microvascular cerebral blood flow in adults at risk for Alzheimer's disease. *Alzheimers Dement (Amst)*, *7*, 48-55. https://doi.org/10.1016/j.dadm.2017.01.002

Cote, S., Butler, R., Michaud, V., Lavallee, E., Croteau, E., Mendrek, A.,···Whittingstall, K. (2021). The regional effect of serum hormone levels on cerebral blood flow in healthy nonpregnant women. *Hum Brain Mapp*, *42*(17), 5677-5688. https://doi.org/10.1002/hbm.25646

Desikan, R. S., Ségonne, F., Fischl, B., Quinn, B. T., Dickerson, B. C., Blacker, D.,⋯Killiany, R. J. (2006). An automated labeling system for subdividing the human cerebral cortex on MRI scans into gyral based regions of interest. *Neuroimage*, *31*(3), 968-980. https://doi.org/10.1016/j.neuroimage.2006.01.021

Detre, J. A., Rao, H., Wang, D. J. J., Chen, Y. F., & Wang, Z. (2012). Applications of arterial spin labeled MRI in the brain. *Journal of Magnetic Resonance Imaging*, *35*(5), 1026-1037. https://doi.org/https://doi.org/10.1002/jmri.23581

Detre, J. A., Wang, J., Wang, Z., & Rao, H. (2009). Arterial spin-labeled perfusion MRI in basic and clinical neuroscience. *Current Opinion in Neurology*, *22*(4). https://journals.lww.com/co-neurology/fulltext/2009/08000/arterial_spin_labeled_perfusion_mri_in_basic_and.4.aspx

Di Biase, Maria A., Tian, Ye E., Bethlehem, Richard A. I., Seidlitz, J., Alexander-Bloch, A. F., Yeo, B. T. T., & Zalesky, A. (2023). Mapping human brain charts cross-sectionally and longitudinally. *Proceedings of the National Academy of Sciences*, *120*(20), e2216798120. https://doi.org/10.1073/pnas.2216798120

Dolui, S., Wang, Z., Wolf, R. L., Nabavizadeh, A., Xie, L., Tosun, D.,⋯Detre, J. A. (2024). Automated Quality Evaluation Index for Arterial Spin Labeling Derived Cerebral Blood Flow Maps. *J Magn Reson Imaging*, *60*(6), 2497-2508. https://doi.org/10.1002/jmri.29308

Dow-Edwards, D., MacMaster, F. P., Peterson, B. S., Niesink, R., Andersen, S., & Braams, B. R. (2019). Experience during adolescence shapes brain development: From synapses and networks to normal and pathological behavior. *Neurotoxicology and Teratology*, *76*, 106834. https://doi.org/https://doi.org/10.1016/j.ntt.2019.106834

Duan, W., Zhou, G. D., Balachandrasekaran, A., Bhumkar, A. B., Boraste, P. B., Becker, J. T.,⋯Dai, W. (2021). Cerebral Blood Flow Predicts Conversion of Mild Cognitive Impairment into Alzheimer's Disease and Cognitive Decline: An Arterial Spin Labeling Follow-up Study. *J Alzheimers Dis*, *82*(1), 293-305. https://doi.org/10.3233/jad-210199

Dukart, J., Holiga, Š., Chatham, C., Hawkins, P., Forsyth, A., McMillan, R.,⋯Sambataro, F. (2018). Cerebral blood flow predicts differential neurotransmitter activity. *Sci Rep*, *8*(1), 4074. https://doi.org/10.1038/s41598-018-22444-0

Fantini, S., Sassaroli, A., Tgavalekos, K. T., & Kornbluth, J. (2016). Cerebral blood flow and autoregulation: current measurement techniques and prospects for noninvasive optical methods. *Neurophotonics*, *3*(3), 031411. https://doi.org/10.1117/1.NPh.3.3.031411

Glasser, M. F., Sotiropoulos, S. N., Wilson, J. A., Coalson, T. S., Fischl, B., Andersson, J. L.,⋯Jenkinson, M. (2013). The minimal preprocessing pipelines for the Human Connectome Project. *Neuroimage*, *80*, 105-124. https://doi.org/10.1016/j.neuroimage.2013.04.127

Hernandez, D. A., Bokkers, R. P., Mirasol, R. V., Luby, M., Henning, E. C., Merino, J. G.,⋯Latour, L. L. (2012). Pseudocontinuous arterial spin labeling quantifies relative cerebral blood flow in acute stroke. *Stroke*, *43*(3), 753-758. https://doi.org/10.1161/strokeaha.111.635979

Hijman, A. S., Wehrle, F. M., Latal, B., Hagmann, C. F., & O'Gorman, R. L. (2024). Cerebral perfusion differences are linked to executive function performance in very preterm-born children and adolescents. *Neuroimage*, *285*, 120500. https://doi.org/10.1016/j.neuroimage.2023.120500

Holz, N. E., Zabihi, M., Kia, S. M., Monninger, M., Aggensteiner, P.-M., Siehl, S.,⋯Consortium, I.

(2023). A stable and replicable neural signature of lifespan adversity in the adult brain. *Nature Neuroscience*, *26*(9), 1603-1612. https://doi.org/10.1038/s41593-023-01410-8

Hu, W. T., Wang, Z., Lee, V. M., Trojanowski, J. Q., Detre, J. A., & Grossman, M. (2010a). Distinct cerebral perfusion patterns in FTLD and AD. *Neurology*, *75*(10), 881-888. https://doi.org/10.1212/WNL.0b013e3181f11e35

Hu, W. T., Wang, Z., Lee, V. M. Y., Trojanowski, J. Q., Detre, J. A., & Grossman, M. (2010b). Distinct cerebral perfusion patterns in FTLD and AD. *Neurology*, *75*(10), 881-888. https://doi.org/10.1212/WNL.0b013e3181f11e35

Kaplan, L., Chow, B. W., & Gu, C. (2020). Neuronal regulation of the blood-brain barrier and neurovascular coupling. *Nat Rev Neurosci*, *21*(8), 416-432. https://doi.org/10.1038/s41583-020-0322-2

Kastrup, A., Kruger, G., Neumann-Haefelin, T., Glover, G. H., & Moseley, M. E. (2002). Changes of cerebral blood flow, oxygenation, and oxidative metabolism during graded motor activation [Research Support, Non-U.S. Gov't

Research Support, U.S. Gov't, P.H.S.]. *NeuroImage*, *15*(1), 74-82. https://doi.org/10.1006/nimg.2001.0916

Kisler, K., Nelson, A. R., Montagne, A., & Zlokovic, B. V. (2017). Cerebral blood flow regulation and neurovascular dysfunction in Alzheimer disease. *Nature Reviews Neuroscience*, *18*(7), 419-434. https://doi.org/10.1038/nrn.2017.48

Krause, D. N., Duckles, S. P., & Pelligrino, D. A. (2006). Influence of sex steroid hormones on cerebrovascular function. *J Appl Physiol (1985)*, *101*(4), 1252-1261. https://doi.org/10.1152/japplphysiol.01095.2005

Kuzawa, C. W., Chugani, H. T., Grossman, L. I., Lipovich, L., Muzik, O., Hof, P. R.,···Lange, N. (2014). Metabolic costs and evolutionary implications of human brain development. *Proceedings of the National Academy of Sciences*, *111*(36), 13010-13015. https://doi.org/10.1073/pnas.1323099111

Leung, J., Kosinski, P. D., Croal, P. L., & Kassner, A. (2016). Developmental trajectories of cerebrovascular reactivity in healthy children and young adults assessed with magnetic resonance imaging. *J Physiol*, *594*(10), 2681-2689. https://doi.org/10.1113/jp271056

Leung, K. K., Bartlett, J. W., Barnes, J., Manning, E. N., Ourselin, S., Fox, N. C., & for the Alzheimer's Disease Neuroimaging, I. (2013). Cerebral atrophy in mild cognitive impairment and Alzheimer disease. *Neurology*, *80*(7), 648-654. https://doi.org/10.1212/WNL.0b013e318281ccd3

Li, X., Hui, Y., Shi, H., Zhao, X., Li, R., Chen, Q.,···Wang, Z. (2023). Association of blood pressure with brain perfusion and structure: A population-based prospective study. *Eur J Radiol*, *165*, 110889. https://doi.org/10.1016/j.ejrad.2023.110889

Li, Y., Dolui, S., Xie, D. F., Wang, Z., & Alzheimer's Disease Neuroimaging, I. (2018). Priors-guided slice-wise adaptive outlier cleaning for arterial spin labeling perfusion MRI. *Journal of Neuroscience Methods*, *307*, 248-253. https://doi.org/10.1016/j.jneumeth.2018.06.007

Li, Y., Sadiq, A., & Wang, Z. (2022). Arterial Spin Labelling-Based Blood-Brain Barrier Assessment and Its Applications. *Investig Magn Reson Imaging*, *26*(4), 229-236. https://doi.org/10.13104/imri.2022.26.4.229

Li, Y., & Wang, Z. (2023). Deeply Accelerated Arterial Spin Labeling Perfusion MRI for Measuring Cerebral Blood Flow and Arterial Transit Time. *IEEE J Biomed Health Inform*, *27*(12), 5937-

5945. https://doi.org/10.1109/jbhi.2023.3312662

Liao, W., Wang, Z., Zhang, X., Shu, H., Wang, Z., Liu, D., & Zhang, Z. (2017). Cerebral blood flow changes in remitted early-and late-onset depression patients. *Oncotarget*, *8*(44), 76214.

Liu, W., Lou, X., & Ma, L. (2016). Use of 3D pseudo-continuous arterial spin labeling to characterize sex and age differences in cerebral blood flow. *Neuroradiology*, *58*(9), 943-948. https://doi.org/10.1007/s00234-016-1713-y

Logothetis, N. K., & Wandell, B. A. (2004). Interpreting the BOLD signal. *Annu Rev Physiol*, *66*, 735-769. https://doi.org/10.1146/annurev.physiol.66.082602.092845

Luis Hernandez-Garcia, V. A., Weiying Dai, Maria A Fernandez-Seara, Jia Guo, Matthias Guenther, Jonas Schollenberger, Ananth J. Madhuranthakam, Henk Mutsaerts, Jan Petr, Qin Qin, Yuriko Suzuki, Manuel Taso, David L. Thomas, Matthias J P van Osch, Joseph G Woods, Moss Y Zhao, Lirong Yan, Ze Wang, Li Zhao, Thomas W Okell, ISMRM Perfusion Study Group. (2022). Recent technical developments in ASL: A Review of the State of the Art. *Magnetic Resonance in Medicine*, *88*(5), 2021-2042. https://doi.org/10.1002/mrm.29381

Mahdi, E. S., Bouyssi-Kobar, M., Jacobs, M. B., Murnick, J., Chang, T., & Limperopoulos, C. (2018). Cerebral Perfusion Is Perturbed by Preterm Birth and Brain Injury. *AJNR Am J Neuroradiol*, *39*(7), 1330-1335. https://doi.org/10.3174/ajnr.A5669

Marquand, A. F., Kia, S. M., Zabihi, M., Wolfers, T., Buitelaar, J. K., & Beckmann, C. F. (2019). Conceptualizing mental disorders as deviations from normative functioning. *Molecular Psychiatry*, *24*(10), 1415-1424. https://doi.org/10.1038/s41380-019-0441-1

Mutsaerts, H., Mirza, S. S., Petr, J., Thomas, D. L., Cash, D. M., Bocchetta, M.,···Masellis, M. (2019). Cerebral perfusion changes in presymptomatic genetic frontotemporal dementia: a GENFI study. *Brain*, *142*(4), 1108-1120. https://doi.org/10.1093/brain/awz039

Nagaraj, U. D., Evangelou, I. E., Donofrio, M. T., Vezina, L. G., McCarter, R., du Plessis, A. J., & Limperopoulos, C. (2015). Impaired Global and Regional Cerebral Perfusion in Newborns with Complex Congenital Heart Disease. *J Pediatr*, *167*(5), 1018-1024. https://doi.org/10.1016/j.jpeds.2015.08.004

Neath, A. A., & Cavanaugh, J. E. (2012). The Bayesian information criterion: background, derivation, and applications. *Wiley Interdisciplinary Reviews: Computational Statistics*, *4*(2), 199-203.

Ngo, A., Royer, J., Rodriguez-Cruces, R., Xie, K., DeKraker, J., Auer, H.,···Bernhardt, B. C. (2024). Associations of Cerebral Blood Flow Patterns With Gray and White Matter Structure in Patients With Temporal Lobe Epilepsy. *Neurology*, *103*(3), e209528. https://doi.org/10.1212/wnl.0000000000209528

Nikiruy, K., Perez, E., Baroni, A., Reddy, K. D. S., Pechmann, S., Wenger, C., & Ziegler, M. (2024). Blooming and pruning: learning from mistakes with memristive synapses. *Sci Rep*, *14*(1), 7802. https://doi.org/10.1038/s41598-024-57660-4

Ouyang, M., Detre, J. A., Hyland, J. L., Sindabizera, K. L., Kuschner, E. S., Edgar, J. C.,···Huang, H. (2024). Spatiotemporal cerebral blood flow dynamics underlies emergence of the limbic-sensorimotor-association cortical gradient in human infancy. *Nature Communications*, *15*(1), 8944. https://doi.org/10.1038/s41467-024-53354-7

Paniukov, D., Lebel, R. M., Giesbrecht, G., & Lebel, C. (2020). Cerebral blood flow increases across early childhood. *Neuroimage*, *204*, 116224. https://doi.org/10.1016/j.neuroimage.2019.116224

Rutherford, S., Barkema, P., Tso, I. F., Sripada, C., Beckmann, C. F., Ruhe, H. G., & Marquand, A. F.

(2023). Evidence for embracing normative modeling. *Elife*, *12*, e85082. https://doi.org/10.7554/eLife.85082

Rutherford, S., Kia, S. M., Wolfers, T., Fraza, C., Zabihi, M., Dinga, R.,⋯Marquand, A. F. (2022). The normative modeling framework for computational psychiatry. *Nat Protoc*, *17*(7), 1711-1734. https://doi.org/10.1038/s41596-022-00696-5

Satterthwaite, T. D., Shinohara, R. T., Wolf, D. H., Hopson, R. D., Elliott, M. A., Vandekar, S. N.,⋯Gur, R. E. (2014). Impact of puberty on the evolution of cerebral perfusion during adolescence. *Proceedings of the National Academy of Sciences*, *111*(23), 8643-8648. https://doi.org/10.1073/pnas.1400178111

Schaeffer, S., & Iadecola, C. (2021). Revisiting the neurovascular unit. *Nat Neurosci*, *24*(9), 1198-1209. https://doi.org/10.1038/s41593-021-00904-7

Schmithorst, V. J., Vannest, J., Lee, G., Hernandez-Garcia, L., Plante, E., Rajagopal, A., & Holland, S. K. (2015). Evidence that neurovascular coupling underlying the BOLD effect increases with age during childhood. *Hum Brain Mapp*, *36*(1), 1-15. https://doi.org/10.1002/hbm.22608

Segal, A., Parkes, L., Aquino, K., Kia, S. M., Wolfers, T., Franke, B.,⋯Fornito, A. (2023). Regional, circuit and network heterogeneity of brain abnormalities in psychiatric disorders. *Nature Neuroscience*, *26*(9), 1613-1629. https://doi.org/10.1038/s41593-023-01404-6

Silbereis, J. C., Pochareddy, S., Zhu, Y., Li, M., & Sestan, N. (2016). The Cellular and Molecular Landscapes of the Developing Human Central Nervous System. *Neuron*, *89*(2), 248-268. https://doi.org/10.1016/j.neuron.2015.12.008

Solis, E., Jr., Hascup, K. N., & Hascup, E. R. (2020). Alzheimer's Disease: The Link Between Amyloid-β and Neurovascular Dysfunction. *J Alzheimers Dis*, *76*(4), 1179-1198. https://doi.org/10.3233/jad-200473

Stasinopoulos, D. M., & Rigby, R. A. (2008). Generalized additive models for location scale and shape (GAMLSS) in R. *Journal of Statistical Software*, *23*, 1-46.

Sun, L., Zhao, T., Liang, X., Xia, M., Li, Q., Liao, X.,⋯He, Y. (2024). Functional connectome through the human life span. *bioRxiv*. https://doi.org/10.1101/2023.09.12.557193

Takahashi, T., Shirane, R., Sato, S., & Yoshimoto, T. (1999). Developmental Changes of Cerebral Blood Flow and Oxygen Metabolism in Children. *American Journal of Neuroradiology*, *20*(5), 917. http://www.ajnr.org/content/20/5/917.abstract

Taylor, J. M., Chang, M., Vaughan, J., Horn, P. S., Zhang, B., Leach, J. L.,⋯Abruzzo, T. (2022). Cerebral Arterial Growth in Childhood. *Pediatr Neurol*, *134*, 59-66. https://doi.org/https://doi.org/10.1016/j.pediatrneurol.2022.06.017

Tierney, A. L., & Nelson, C. A., 3rd. (2009). Brain Development and the Role of Experience in the Early Years. *Zero Three*, *30*(2), 9-13.

Tsujikawa, T., Kimura, H., Matsuda, T., Fujiwara, Y., Isozaki, M., Kikuta, K., & Okazawa, H. (2016). Arterial Transit Time Mapping Obtained by Pulsed Continuous 3D ASL Imaging with Multiple Post-Label Delay Acquisitions: Comparative Study with PET-CBF in Patients with Chronic Occlusive Cerebrovascular Disease. *PLoS One*, *11*(6), e0156005. https://doi.org/10.1371/journal.pone.0156005

van Dinther, M., Hooghiemstra, A. M., Bron, E. E., Versteeg, A., Leeuwis, A. E., Kalay, T.,⋯van Oostenbrugge, R. J. (2024). Lower cerebral blood flow predicts cognitive decline in patients with vascular cognitive impairment. *Alzheimers Dement*, *20*(1), 136-144. https://doi.org/10.1002/alz.13408

Vandekar, S. N., Shou, H., Satterthwaite, T. D., Shinohara, R. T., Merikangas, A. K., Roalf, D. R.,⋯Detre, J. A. (2017). Sex differences in estimated brain metabolism in relation to body growth through adolescence. *Journal of Cerebral Blood Flow & Metabolism*, *39*(3), 524-535. https://doi.org/10.1177/0271678X17737692

Verdi, S., Kia, S. M., Yong, K. X. X., Tosun, D., Schott, J. M., Marquand, A. F., & Cole, J. H. (2023). Revealing Individual Neuroanatomical Heterogeneity in Alzheimer Disease Using Neuroanatomical Normative Modeling. *Neurology*, *100*(24), e2442-e2453. https://doi.org/10.1212/WNL.0000000000207298

Verdi, S., Rutherford, S., Fraza, C., Tosun, D., Altmann, A., Raket, L. L.,⋯for the Alzheimer's Disease Neuroimaging, I. (2024). Personalizing progressive changes to brain structure in Alzheimer's disease using normative modeling. *Alzheimer's & Dementia*, *20*(10), 6998-7012. https://doi.org/https://doi.org/10.1002/alz.14174

Vidorreta, M., Balteau, E., Wang, Z., De Vita, E., Pastor, M. A., Thomas, D. L.,⋯Fernandez-Seara, M. A. (2014). Evaluation of segmented 3D acquisition schemes for whole-brain high-resolution arterial spin labeling at 3 T. *NMR in Biomedicine*, *27*(11), 1387-1396. https://doi.org/10.1002/nbm.3201

Vidorreta, M., Wang, Z., Chang, Y. V., Wolk, D. A., Fernández-Seara, M. A., & Detre, J. A. (2017). Whole-brain background-suppressed pCASL MRI with 1D-accelerated 3D RARE Stack-Of-Spirals readout. *PLoS One*, *12*(8), e0183762. https://doi.org/10.1371/journal.pone.0183762

Vidorreta, M., Wang, Z., Rodriguez, I., Pastor, M. A., Detre, J. A., & Fernandez-Seara, M. A. (2012). Comparison of 2D and 3D single-shot ASL perfusion fMRI sequences. *NeuroImage*, *66C*, 662-671. https://doi.org/10.1016/j.neuroimage.2012.10.087

Wang, Z. (2012). Improving Cerebral Blood Flow Quantification for Arterial Spin Labeled Perfusion MRI by Removing Residual Motion Artifacts and Global Signal Fluctuations. *Magnetic Resonance Imaging*, *30*(10), 1409-1415. https://doi.org/http://dx.doi.org/10.1016/j.mri.2012.05.004

Wang, Z. (2014). Characterizing Early Alzheimer's Disease and Disease Progression Using Hippocampal Volume and Arterial Spin Labeling Perfusion MRI. *Journal of Alzheimer's disease*, *42*, S495-S502. https://doi.org/Doi 10.3233/Jad-141419

Wang, Z. (2022). Arterial Spin Labeling Perfusion MRI Signal Processing Through Traditional Methods and Machine Learning. *Investigative Magnetic Resoance Imaging*, *26*(4), 220-228. https://doi.org/ https://doi.org/10.13104/imri.2022.26.4.220

Wang, Z. (2022). Arterial Spin Labeling Perfusion MRI Signal Processing Through Traditional Methods and Machine Learning. *Investig Magn Reson Imaging*, *26*(4), 220-228. https://doi.org/10.13104/imri.2022.26.4.220

Wang, Z., Aguirre, G. K., Rao, H., Wang, J., Fernández-Seara, M. A., Childress, A. R., & Detre, J. A. (2008). Empirical optimization of ASL data analysis using an ASL data processing toolbox: ASLtbx. *Magn Reson Imaging*, *26*(2), 261-269. https://doi.org/10.1016/j.mri.2007.07.003

Wang, Z., Das, S. R., Xie, S. X., Arnold, S. E., Detre, J. A., & Wolk, D. A. (2013). Arterial spin labeled MRI in prodromal Alzheimer's disease: A multi-site study. *NeuroImage: Clinical*, *2*, 630-636. https://doi.org/https://doi.org/10.1016/j.nicl.2013.04.014

Wang, Z., Faith, M., Patterson, F., Tang, K., Kerrin, K., Wileyto, E. P.,⋯Lerman, C. (2007). Neural substrates of abstinence-induced cigarette cravings in chronic smokers. *J Neurosci*, *27*(51),

14035-14040, PMC2153440. https://doi.org/27/51/14035 [pii] 10.1523/JNEUROSCI.2966-07.2007

Wang, Z., Fernandez-Seara, M., Alsop, D. C., Liu, W. C., Flax, J. F., Benasich, A. A., & Detre, J. A. (2008). Assessment of functional development in normal infant brain using arterial spin labeled perfusion MRI. *NeuroImage*, *39*(3), 973-978. https://doi.org/10.1016/j.neuroimage.2007.09.045

Wei, W., Karim, H. T., Lin, C., Mizuno, A., Andreescu, C., Karp, J. F.,···Aizenstein, H. J. (2018). Trajectories in Cerebral Blood Flow Following Antidepressant Treatment in Late-Life Depression: Support for the Vascular Depression Hypothesis. *J Clin Psychiatry*, *79*(6). https://doi.org/10.4088/JCP.18m12106

Wolfers, T., Doan, N. T., Kaufmann, T., Alnæs, D., Moberget, T., Agartz, I.,···Marquand, A. F. (2018). Mapping the Heterogeneous Phenotype of Schizophrenia and Bipolar Disorder Using Normative Models. *JAMA Psychiatry*, *75*(11), 1146-1155. https://doi.org/10.1001/jamapsychiatry.2018.2467

Wolters, F. J., Zonneveld, H. I., Hofman, A., van der Lugt, A., Koudstaal, P. J., Vernooij, M. W., & Ikram, M. A. (2017). Cerebral Perfusion and the Risk of Dementia: A Population-Based Study. *Circulation*, *136*(8), 719-728. https://doi.org/10.1161/circulationaha.117.027448

Wong, A. M., Yan, F.-X., & Liu, H.-L. (2014). Comparison of three-dimensional pseudo-continuous arterial spin labeling perfusion imaging with gradient-echo and spin-echo dynamic susceptibility contrast MRI. *Journal of Magnetic Resonance Imaging*, *39*(2), 427-433. https://doi.org/https://doi.org/10.1002/jmri.24178

Wu, C., Honarmand, A. R., Schnell, S., Kuhn, R., Schoeneman, S. E., Ansari, S. A.,···Shaibani, A. (2016). Age-Related Changes of Normal Cerebral and Cardiac Blood Flow in Children and Adults Aged 7 Months to 61 Years. *J Am Heart Assoc*, *5*(1). https://doi.org/10.1161/jaha.115.002657

Xu, G., Rowley, H. A., Wu, G., Alsop, D. C., Shankaranarayanan, A., Dowling, M.,···Johnson, S. C. (2010). Reliability and precision of pseudo-continuous arterial spin labeling perfusion MRI on 3.0 T and comparison with 15O-water PET in elderly subjects at risk for Alzheimer's disease [Comparative Study

Research Support, N.I.H., Extramural

Research Support, U.S. Gov't, Non-P.H.S.]. *NMR in Biomedicine*, *23*(3), 286-293. https://doi.org/10.1002/nbm.1462

Yeo, B. T., Krienen, F. M., Sepulcre, J., Sabuncu, M. R., Lashkari, D., Hollinshead, M.,···Buckner, R. L. (2011). The organization of the human cerebral cortex estimated by intrinsic functional connectivity. *J Neurophysiol*, *106*(3), 1125-1165. https://doi.org/10.1152/jn.00338.2011

Yu, Q., Ouyang, M., Detre, J., Kang, H., Hu, D., Hong, B.,···Huang, H. (2023). Infant brain regional cerebral blood flow increases supporting emergence of the default-mode network. *Elife*, *12*. https://doi.org/10.7554/eLife.78397

Ze Wang, J. S., Dingna Duan, Stefanie Darnley, Ying Jing, Jian Zhang, Charles O'Brien, Anna Rose Childress. (2017). A Hypo-Status in Drug Dependent Brain Revealed by Multi-modal MRI. *Addiction Biology*, *22*(6), 1622-1631. http://www.ncbi.nlm.nih.gov/pubmed/25957794

Zhang, N., Gordon, M. L., Ma, Y., Chi, B., Gomar, J. J., Peng, S.,···Goldberg, T. E. (2018). The Age-Related Perfusion Pattern Measured With Arterial Spin Labeling MRI in Healthy Subjects. *Front Aging Neurosci*, *10*, 214. https://doi.org/10.3389/fnagi.2018.00214